\documentclass[a4paper]{spie}  

\usepackage{amsmath,amsfonts,amssymb}
\usepackage{graphicx}
\usepackage[colorlinks=true, allcolors=blue]{hyperref}

\def\spose#1{\hbox to 0pt{#1\hss}}
\def\simlt{\mathrel{\spose{\lower 3pt\hbox{$\mathchar"218$}}
     \raise 2.0pt\hbox{$\mathchar"13C$}}}
\def\simgt{\mathrel{\spose{\lower 3pt\hbox{$\mathchar"218$}}
     \raise 2.0pt\hbox{$\mathchar"13E$}}}

\def\arcsec{\nobreak{$''$}}

\title{The MICADO first light imager for the ELT: overview, operation, simulation}

\author[a]{R.~Davies}
\author[b]{J.~Alves}
\author[c]{Y.~Cl\'enet}
\author[d]{F.~Lang-Bardl}
\author[e]{H.~Nicklas}
\author[f]{J.-U.~Pott}
\author[g]{R.~Ragazzoni}
\author[h]{E.~Tolstoy}
\author[i]{P.~Amico}
\author[e]{H.~Anwand-Heerwart}
% ...
\affil[a]{Max Planck Institute for extraterrestrial Physics, 85748 Garching, Germany}
\affil[b]{Institute for Astrophysics, University of Vienna, 1180 Wien, Austria}
\affil[c]{LESIA, Observatoire de Paris, Universit\'e PSL, CNRS, 92195, Meudon, France}
\affil[d]{University Observatory of Munich, 81679 M\"unchen, Germany}
\affil[e]{Institute for Astrophysics, Georg-August-Universit\"at G\"ottingen, 37077 G\"ottingen, Germany}
\affil[f]{Max Planck Institute for Astronomy, 69117 Heidelberg, Germany}
\affil[g]{INAF -- Osservatorio Astronomico di Padova, 35122, Padova, Italy}
\affil[h]{Kapteyn Astronomical Institute, 9700 AV Groningen, The Netherlands}
\affil[i]{European Southern Observatory, 85748, Garching, Germany}
\affil[j]{Leiden Observatory, University of Leiden, 2300 RA Leiden, The Netherlands}
\affil[k]{GEPI, Observatoire de Paris, Universit\'e PSL, CNRS, 92195, Meudon, France}
\affil[l]{Institute for System Dynamics, University of Stuttgart, 70563 Stuttgart, Germany}
\affil[m]{Industrial Mathematics Institute, Johannes Kepler University Linz, 4040 Linz, Austria}
\affil[n]{NOVA-ASTRON, Oude Hoogeveensedijk, 7991 PD Dwingeloo, The Netherlands}
\affil[o]{Institut f\"ur Astro- und Teilchenphysik, Universit\"at Innsbruck, 6020 Innsbruck, Austria}
\affil[p]{LERMA, Observatoire de Paris, Universit\'e PSL, CNRS, 75014 Paris, France}
\affil[q]{Johann Radon Institute for Computational and Applied Mathematics, 4040 Linz, Austria}

\authorinfo{Further author information: send correspondence to R. Davies, email davies@mpe.mpg.de}

\begin{document} 
\maketitle

\begin{abstract}
MICADO will enable the ELT to perform diffraction limited near-infrared observations at first light. 
The instrument's capabilities focus on imaging (including astrometric and high contrast) as well as single object spectroscopy. 
This contribution looks at how requirements from the observing modes have driven the instrument design and functionality. 
Using examples from specific science cases, and making use of the data simulation tool, an outline is presented of what we can expect the instrument to achieve.
\end{abstract}

\keywords{Adaptive Optics, Near-infrared, Cryogenic, Imaging, Astrometry, High Contrast, Spectroscopy, ELT}

\section{Introduction}
\label{sec:intro}  

MICADO\cite{dav16}, the Multi-AO Imaging Camera for Deep Observations, will equip the ELT with a first light diffraction limited imaging capability at near-infrared wavelengths. 
The instrument will work with a multi-conjugate laser guide star adaptive optics system (MCAO, developed by the MAORY consortium\cite{dio16,cil18}) as well as a single-conjugate natural guide star adaptive optics system (SCAO, developed jointly by the MICADO and MAORY consortia\cite{cle18}). 
It will interface to the MAORY warm optical relay that re-images the telescope focus.
In this configuration, both MCAO and SCAO are available.
If required for an initial phase, MICADO will also be able to operate with just the SCAO system in a `stand-alone' mode, using a simpler optical relay that interfaces directly to the telescope.

MICADO has the potential to address a large number of science topics that span the key elements of modern astrophysics. 
The science drivers\cite{tol18} focus on several main themes: the dynamics of dense stellar systems, black holes in galaxies and the centre of the Milky Way, the star formation history of galaxies through resolved stellar populations, the formation and evolution of galaxies in the early universe, planets and planet formation, and the solar system. 
To address these, MICADO will exploit its key capabilities of sensitivity and resolution, which are in turn leveraged by its observing modes of imaging, astrometry, coronagraphy, and spectroscopy.
The main characteristics of these four observing modes are described in Sections~\ref{sec:stdim}--\ref{sec:spectro} below, which show how they have shaped the instrument design and operational concept.
An illustration of the astrophysics that might be addressed is enabled using the instrument data simulator SimCADO\cite{les16}, which is available at http://www.univie.ac.at/simcado.

\section{Overview}
\label{sec:overview}

\begin{figure}
\begin{center}
\includegraphics[width=17cm]{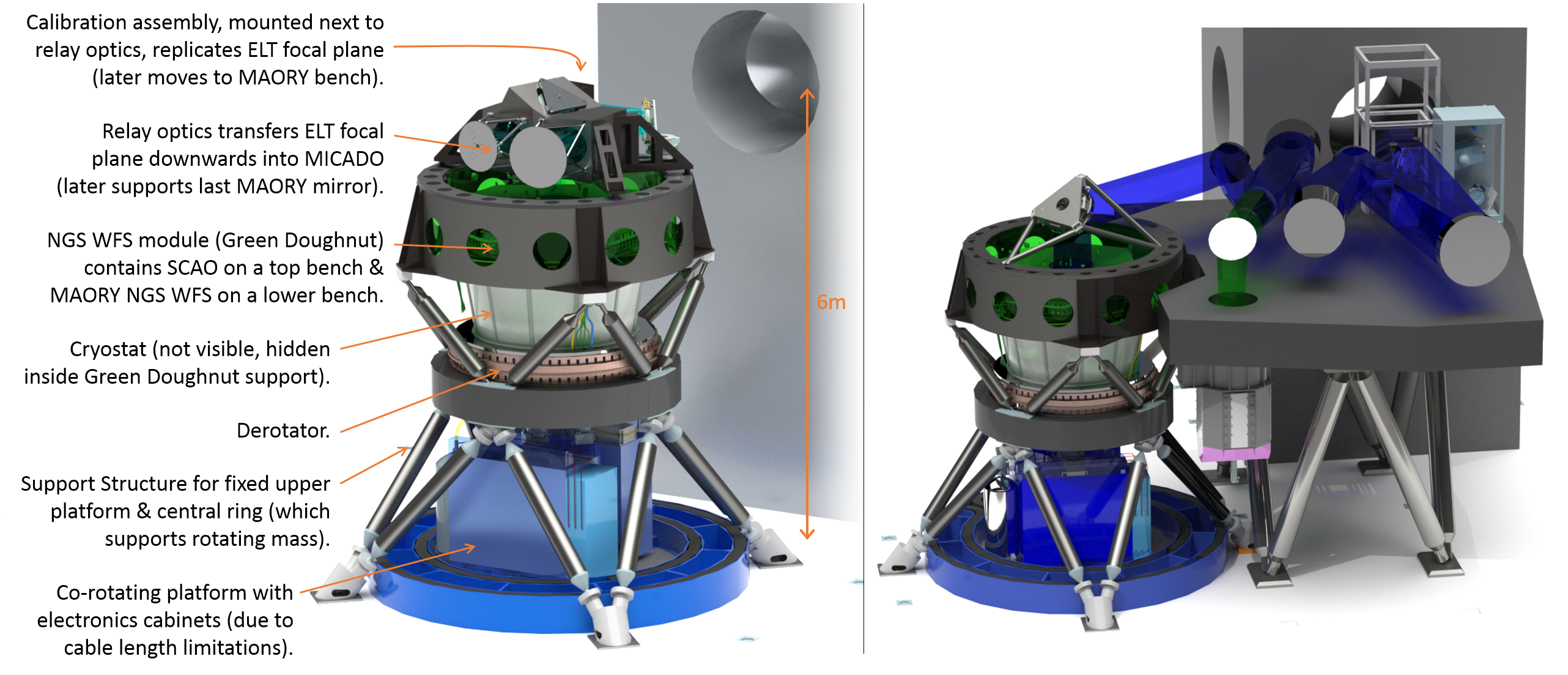}
\end{center}
\caption{Global MICADO architecture. Left: in `stand-alone' mode, with the various sub-systems labelled. Right: with MAORY. In the latter configuration, the stand-alone relay optics are replaced by the last fold mirror in the MAORY optical relay; and the calibration unit is moved to the elevator at the ELT telescope focal plane.}
\label{fig:overview}
\end{figure}

Fig.~\ref{fig:overview} presents a top level summary of the global architecture of MICADO, indicating the various sub-systems.
This is not discussed further here because the system and sub-systems have been described previously\cite{dav16}, and the general concepts have not changed dramatically.
In addition, recent detailed descriptions are given elsewhere for a number of sub-systems and topics: 
the calibration assembly\cite{rod18mca};
the SCAO system\cite{cle18};
the cryostat and cold optics instrument\cite{sch18}, including the main selection mechanism\cite{lan18}, the central wheel mechanism\cite{tro18}, and a test cryostat\cite{mon18};
the derotator\cite{bar18,glu18};
the support structure\cite{nic18};
the instrument control approach\cite{hau18}; 
performance optimisation\cite{gar18};
and PSF reconstruction\cite{wag18}.

\section{Standard Imaging}
\label{sec:stdim}

Standard imaging is the simplest observing mode of MICADO. 
It is designed to obtain images with a diffraction limited resolution at wavelengths in the range 0.8--2.4\,$\mu$m, given that the wavefront error delivered to the instrument by MCAO will be in the range 280--380\,nm across the field, corresponding to 30--50\% Strehl ratio in the K-band. 
With bright stars, SCAO will deliver wavefront errors as low as 200--250\,nm, equivalent to 60--70\% Strehl ratio in K-band on axis.

\subsection{Characteristics}

This mode drives the majority of the basic requirements for MICADO.
Its properties are defined primarily by the cold opto-mechanics inside the cryostat, which are described in detail elsewhere\cite{sch18}.
Key elements are:
\begin{itemize}
\item
Gravity invariant rotation to minimize flexure.
\item
A high throughput and low wavefront error, to optimise the instrument sensitivity. To achieve this, challenging performance requirements have been put on all optical surfaces. Given the wavefront error expected from the AO systems, and the difficulty of correcting non-common path aberrations across the whole field, the optical performance budget sets the goal of reaching only 120\,nm rms wavefront error between the cryostat entrance window and detectors.
\item
An array of 3$\times$3 H4RG detectors, which are read at a pixel rate of $\sim$200\,kHz to allow the most sensitive exposures.
A low resolution imager offers a 4\,mas pixel scale over a 50.5\arcsec$\times$50.5\arcsec\ field of view, to fully
sample the diffraction limited PSF from 1.5\,$\mu$m to 2.4\,$\mu$m; the wide field effectively provides a major multiplex advantage.
A high resolution imager provides a 1.5\,mas pixel scale over a 19\arcsec$\times$19\arcsec\ field of view, to fully sample the diffraction limited PSF from 0.8\,$\mu$m to 1.5\,$\mu$m; the fine sampling at long
wavelengths provides the capability needed for PSF de-blending in very crowded fields.
\item
Large filter wheels, able to hold $>$30 filters, for broad-band and narrow-band imaging as required by the science cases. 
The filter suite includes neutral density filters for bright targets, and spectroscopic filters that could in principle also be used for imaging. 
The large 14\,cm filter diameter, and the requirement to have high throughput (exceeding 95\% for the broad-band filters) as well as steep bandpass edges, will mean the manufacturing is not straightforward. 
Given that for typical observatories, more than $\sim$90\% of observations are performed with only $\sim$10 filters, if a balance is needed between performance, cost, and number of filters, the preference will be for fewer high quality filters. 
The number of available slots, however, would remain the same.
\item
An atmospheric dispersion corrector (ADC), which is always in the beam path.
The chromatic dispersion is large enough with respect to the diffraction limit, that, over nearly the whole sky, the gain in sensitivity by correcting it and keeping the PSF compact outweighs the loss of throughput (due to the ADC's 8 optical surfaces) even for isolated point sources.
\item
Dithering is possible within 0.3\arcsec\ of any pointing (with a small time overhead of $\sim$2\,s), to address detector systematics;
and within $\sim$10\arcsec\ of the initial pointing (with an overhead up to $\sim$15\,s) to derive sky background.
Larger dithers may be possible depending on the NGS configuration used by MCAO.
Offsetting by several arcminutes to sky (without AO correction) is also possible with a somewhat longer $\sim$30\,s overhead.
\item
Secondary guiding is the use of reference sources in the science frames to measure slow drifts of position, rotation, and plate scale (on timescales of seconds to minutes). 
Optionally, these corrections can be applied during observations. 
Information about the reference sources is available to the pipeline, which can use them for fine matching of individual exposures before combining them.
\item
Differential tracking, needed for observing solar system targets that have a non-sidereal speed up to 100\arcsec\,hr$^{-1}$, will be possible with both SCAO and MCAO.
\item
The preparation tool (PreCADO\cite{weg18}) is essential for configuring observations. 
Among its many tasks are: the assessment of how the NGS asterism restricts the dither pattern, and whether some pointings might be allowed where only 2 rather than all 3 NGS are accessible; the identification and handling of `spare' NGS in case any of the primary set cannot be used; notification about sources likely to saturate for the selected exposure time; and estimation of the observing efficiency.
\item
The data pipeline will provide standard processing of the exposures, combining frames from a single observing block to create a photometrically calibrated image, together with an estimate of the noise.
The alignment of the individual exposures will be done to a few milliarcsecond precision. 
While photometric calibration will ideally be done using sources in the science frames,  the distortion correction will rely as little as possible on those. Instead the aim is primarily to use corrections that are pre-calibrated or derived from models.
\end{itemize}

\subsection{Astrophysical Applications}

\begin{figure}
\begin{center}
\includegraphics[width=17cm]{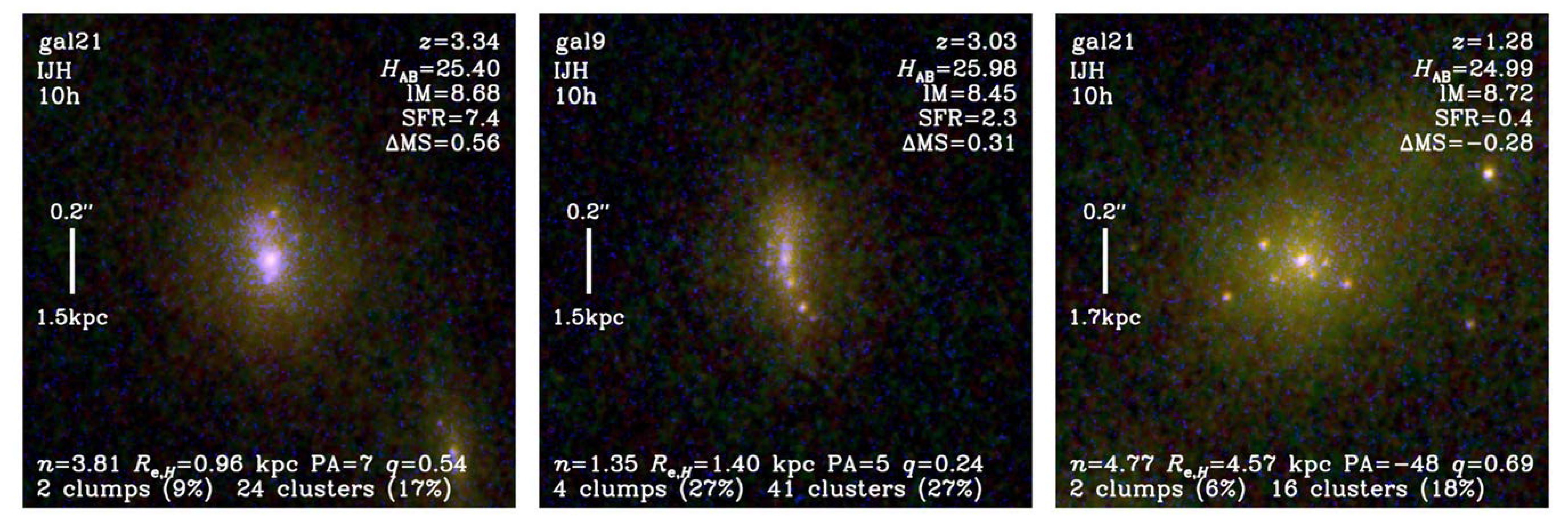}
\end{center}
\caption{Simulations of galaxies above and below the main sequence at $z\sim2$, created with SimCADO. These composite IJH colour maps are based on known galaxies in the Hubble Ultra Deep Field, to which additional inferred structure (in particular particular a star cluster and clump population) has been added. Adapted from the MICADO Science Case\cite{tol18}.}
\label{fig:highz}
\end{figure}

With a point-source sensitivity that is comparable to JWST and a resolution about a factor 6 better, MICADO is well suited to numerous science cases.
One important topic is galaxy evolution over cosmic time.
We now have a fairly robust outline of this evolution for global galaxy properties, and hence the first pieces of evidence about how galaxies assembled and transformed into the present day Hubble sequence.
An obvious next step is to resolve the faint distant galaxies on sufficiently small scales to assess their sub-galactic components including disk structures, nascent bulges, clumps, and globular cluster progenitors.
The current view is limited by spatial resolution, which corresponds to $\sim$1\,kpc in the best cases\cite{for11,for18} (space-based telescopes or adaptive optics on 8-m class ground-based telescopes).
In particular, relatively unexplored regimes include lower mass galaxies, comprising the bulk (by number) of the galaxy population, and galaxies at early cosmic times, when they were building their first stars.
Fig.~\ref{fig:highz} illustrates the type of detailed structure within high redshift galaxies that MICADO might be able to detect.

\begin{figure}
\begin{center}
\includegraphics[width=17cm]{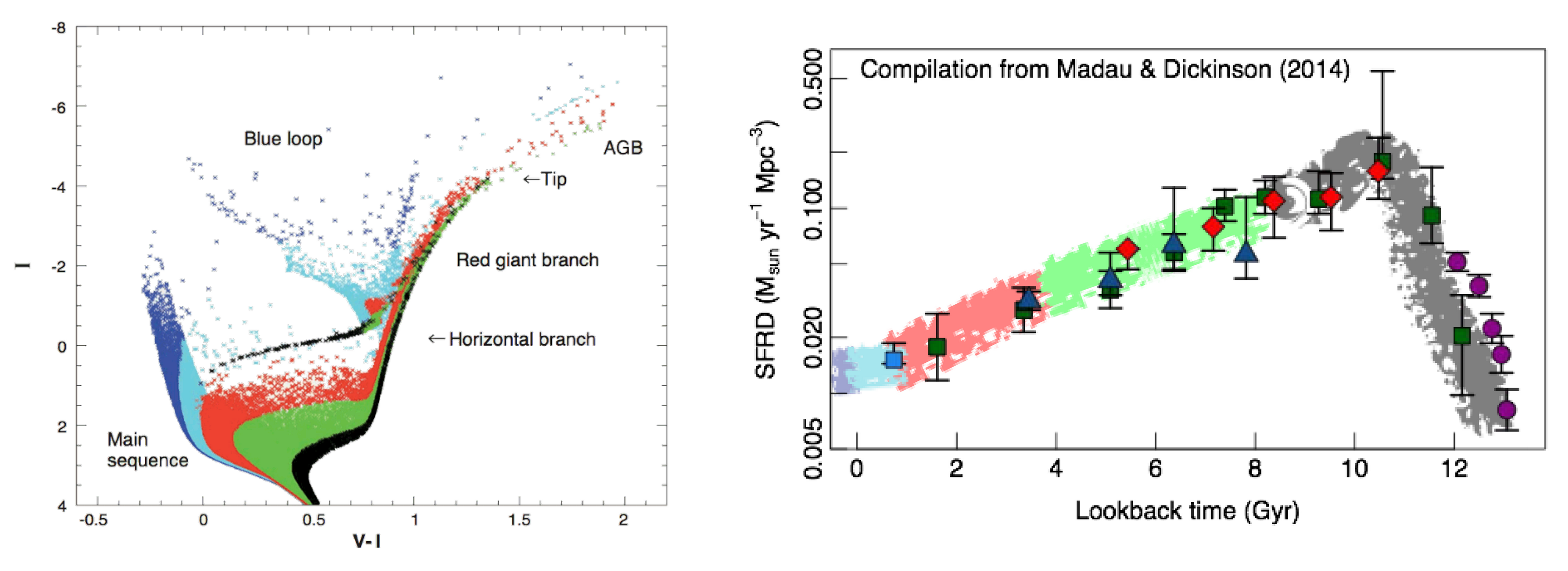}
\end{center}
\caption{Star formation at early times can be probed either by observing high redshift galaxies directly, or via colour magnitude diagrams (CMDs) of the relic stellar populations in local galaxies. The features in the CMD\cite{tol11} (left) are coloured to approximately match those in the plot of cosmic star formation rate density\cite{mad14} (right), to indicate how they are related. Note, however, that MICADO will measure near-infrared CMDs rather than the optical one shown here. Adapted from the MICADO Science Case\cite{tol18}.}
\label{fig:cmd_sfrd}
\end{figure}

\begin{figure}
\begin{center}
\includegraphics[width=17cm]{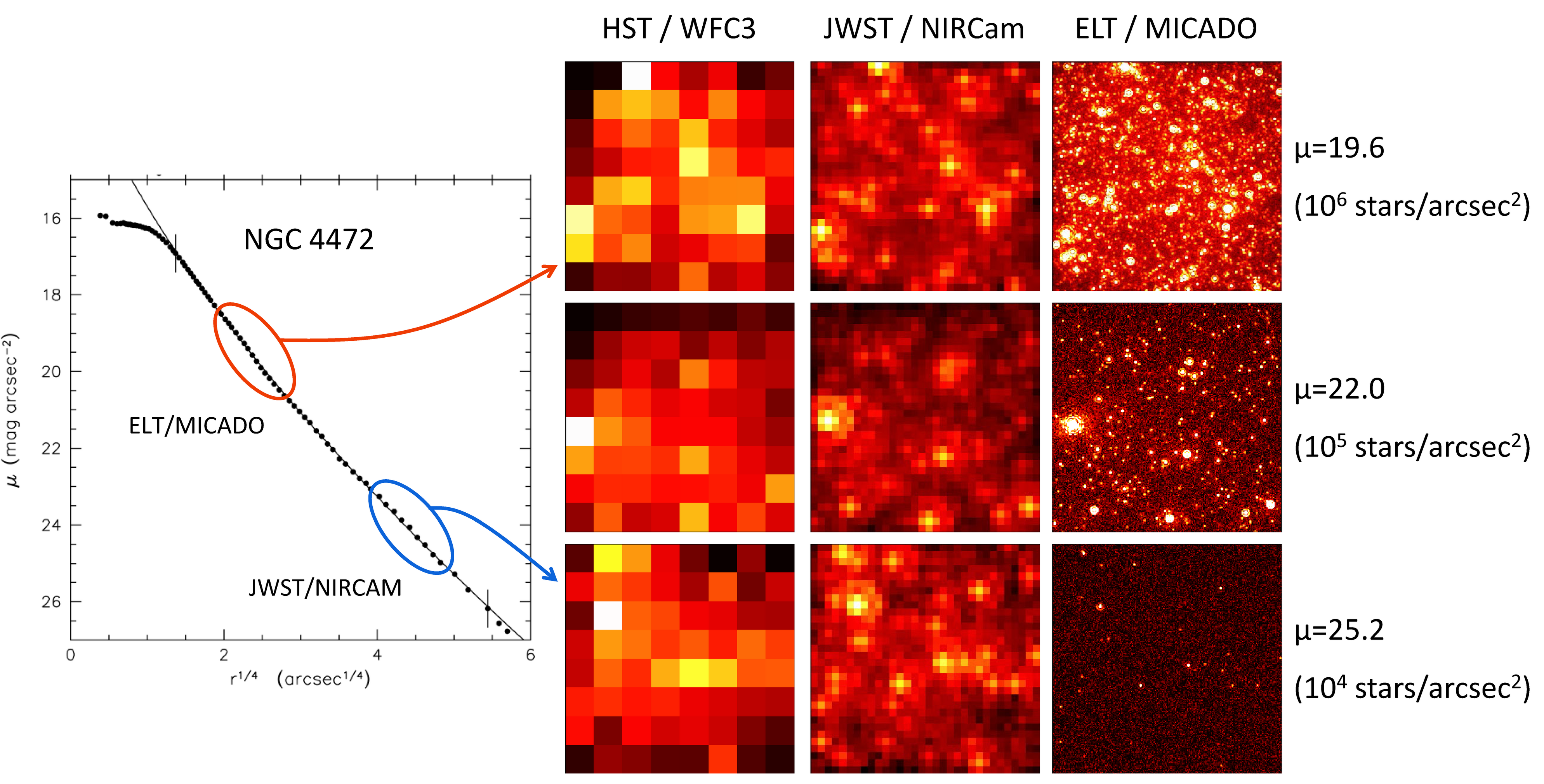}
\end{center}
\caption{Left: surface brightness of NGC\,4472 in the Virgo Cluster, marking the regimes that JWST and MICADO will be able to probe. Right: SimCADO simulations of crowded stellar fields at various surface brightnesses, showing the impact of crowding on what can be measured by HST, JWST, and MICADO.}
\label{fig:stellarpops}
\end{figure}

An alternative probe of galaxy evolution is via the relic populations in local galaxies, by performing photometry on individual stars to generate a colour magnitude diagram (CMD).
Fig.~\ref{fig:cmd_sfrd} demonstrates the way in which stars formed at different cosmic times are related to the various features of a CMD.
Detecting stars on the horizontal branch enables one to trace the star formation history of galaxies to $z > 6$, to the reionization epoch.
The ultimate goal for resolved stellar populations is to probe the central regions of elliptical galaxies in the Virgo Cluster.
The high surface brightness, due to extreme stellar crowding, makes this very challenging.
Fig.~\ref{fig:stellarpops} shows how the surface brightness depends on radius for one such typical galaxy NGC\,4472 (M\,49), together with a simulation showing the impact of crowding at various surface brightnesses.
JWST will only be able to probe the outskirts of these galaxies, while the higher resolution of MICADO will enable it to reach almost to the centre.

\section{Astrometric Imaging}
\label{sec:astrom}

One of the most challenging requirements for MICADO is to perform astrometry over the MCAO corrected field, and reach a precision better than 50\,$\mu$as.
To fulfil this would be a remarkable achievement, since it is a factor 5--10 better than ground-based 8-m class telescopes with current MCAO systems\cite{nei14,lu14,mas16}, a factor 5 better than space telescopes such as HST\cite{and06}, and is similar to that reachable with dedicated astrometry space missions such as Gaia\cite{lin16}.
Developing a methodology for doing so on the ELT, which is not designed to be an astrometric telescope, and in which every mirror is shifting with respect to the others, is a long and difficult process\cite{rod18ast,pot18}.
In addition, studies by the MAORY consortium indicate that positioning of the low order NGS WFS probes, and how they handle distortions from the atmosphere and warm optical relay, is critical to achieving the astrometric requirement.
As such, this requirement can only be met as a collaborative effort between the telescope, AO system, and instrument designers.

In order to address this requirement it is important to distinguish between absolute and relative astrometry.
Absolute astrometry is required when comparing the position of objects observed with different instruments, often in different wavebands.
Examples could include the position of bright spots in the jet of an active galaxy measured at 2\,$\mu$m by MICADO to those measured at 3\,mm from other facilities; or the location of SiO masers at mm wavelengths around a star, the photo-centre of which is measured in the near-infrared light.
This requires a reference frame, and the astrometric precision will usually be limited by the options available for that.
Since this is beyond the control of the instrument design or operation, it can only be done on a best effort basis.
Relative astrometry (or differential absolute astrometry\cite{pot18}), is about changes in position between epochs: proper motions rather than the position itself.
To achieve this, stability and calibration are more important than minimising distortions {\it per se}.
And to make the problem tractable, it is necessary to assess linear distortions separately to second and third order distortions, which are again separated from high order distortions.
This is equivalent to distinguishing spatial scales. 
The three regimes above correspond to the full field, scales of $\sim$10\arcsec, and scales $<$1\arcsec.
A more detailed discussion and analysis is given elsewhere\cite{pot18}.

\subsection{Characteristics}

The aim of this mode is to reach signal-to-noise limited astrometric precision, which is in the range 10--50\,$\mu$as for bright sources.
To achieve this requires a plate scale precision of $10^{-5}$.
Locally, this corresponds to 10\,$\mu$as over a scale of 1\arcsec.
Globally, the precision will be limited by the availability of reference sources.

The characteristics of this observing mode from the design and operation perspective include:
\begin{itemize}
\item
Mechanical and thermal flexure are minimized by: 
(i) the gravity invariant instrument orientation; 
(ii) an optical path in which all mirrors are fixed (for the high resolution imager, the ADC and filters are the only movable optics; the only additional movable optics in the low resolution imager are two flat fold
mirrors); and 
(iii) controlling the temperature of the cold opto-mechanics to $\sim$0.1\,K over a 1-hr observing block.
\item
Calibration masks will be used to measure instrument distortion. 
These will include the distortion both in MICADO and in the warm optical relay, as a function of rotation (for field versus pupil angle); 
and distortions from the ADC as a function of rotation angle (for zenith distance). 
Stabilisation of the temperature in the cryostat will ensure that calibrations are applicable over sufficiently long timescales.
\item
In contrast to standard imaging, observational constraints may need to be imposed. These could include: 
(i) fewer filters may be offered (for example, one broad and one narrow filter for each of the J, H, and K-bands), to avoid excessive calibration load; 
(ii) night-time calibrations may be attached to the science observations (e.g. a distortion measurement at the start and/or end of an observing block);
(iii) specific dither patterns may be used to allow on-the-fly calibration.
(iv) the airmass range may be restricted (not too large, since atmopsheric effects become severe; but not too small, since field rotation increases).  
\item
A reference frame is needed to set the global plate scale and low order distortion terms. 
The Gaia (and perhaps Euclid) reference frame will be an option for this in some cases -- but not all, because many of the fields observed by MICADO will be either crowded or obscured. 
It may be possible to omit a reference frame if the purpose of the observations is to measure only proper motions, rather than positions, of the stars.
\item
The pipeline will process and combine the frames from a single observing block, in such a way as to yield a photometrically and astrometrically calibrated data product. 
Correcting variations, or evolution of, low order distortions between individual frames using astrophysical sources in the data themselves is central to the astrometric data processing. 
This will be possible, because the pointings where the most precise astrometry is required are likely to be stellar fields.
\end{itemize}

\subsection{Astrophysical Applications}

\begin{figure}
\begin{center}
\includegraphics[width=15cm]{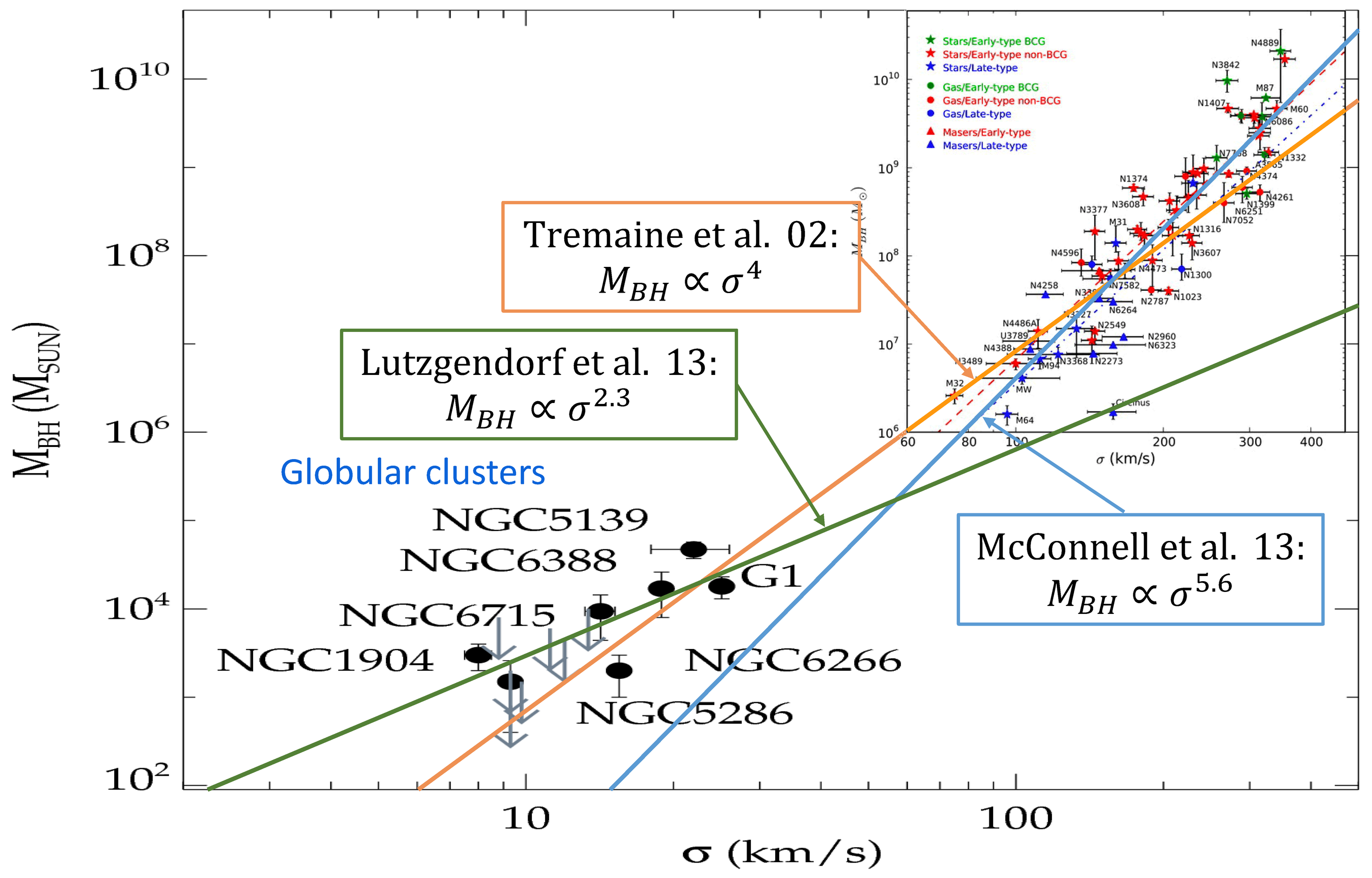}
\end{center}
\caption{Relation between the mass $M_{BH}$ of the central black hole and the velocity dispersion $\sigma$ of the stellar spheroid around it\cite{lut13,tre02,mcc13}. The shallower (green) slope of the relation for globular clusters may suggest different physical processes at work than for galaxies. But note that nearly all of the globular cluster data points (even those drawn as points with errorbars) have $M_{BH}$ estimated at $<$3$\sigma$ significance, or some other systematic issue affecting the measurement. As such, they could be over-estimates, and may actually lie on either of the two relations (orange, blue) measured for galaxies.}
\label{fig:msigma}
\end{figure}

\begin{figure}
\begin{center}
\includegraphics[width=15cm]{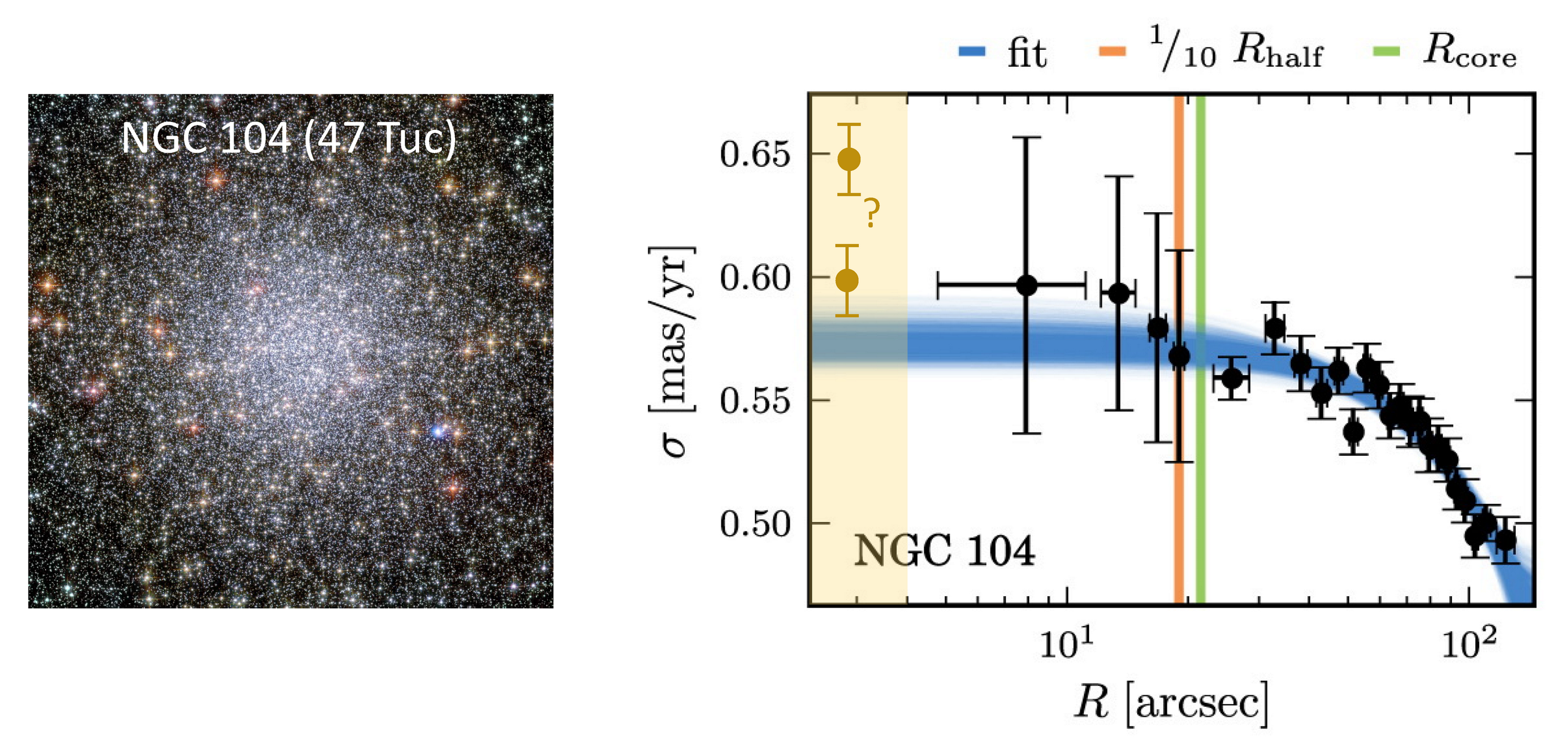}
\end{center}
\caption{Left: image of NGC\,104 (47\,Tuc) from HST. Right: state-of-the-art plot of the radial profile of its stellar velocity dispersion\cite{wat15}. The blue band indicates the range of polynomial fits, forced to be flat at the centre. Crowding limits measurements at radii smaller than the black data points shown. The yellow band indicates the region that must be probed to robustly estimate a black hole mass in this globular cluster, and the yellow points within that region show possible measurements -- with realistic errorbars -- that can be made by MICADO. Adapted from MICADO Science Case\cite{tol18}.}
\label{fig:47tuc}
\end{figure}

One of the immediately obvious rationales for astrometry is using stellar proper motions to probe the existence and masses of black holes in stellar clusters, as well as nearby low mass dwarf galaxies.
There has been an increasing effort in this direction with a variety of tantalising results, but without robust detections\cite{lut13}.
Astrophysically, and as shown in Fig.~\ref{fig:msigma}, one of the key questions concerns the slope of the $M_{BH}-\sigma$ relation between the mass of the central black hole and the velocity dispersion of the stellar spheroid around it.
Initial measurements\cite{tre02} of black holes in elliptical galaxies and classical bulges of disk galaxies had suggested $M_{BH}\propto\sigma^{4}$.
More recent assessments\cite{mcc13} have argued in favour of a steeper slope of 5.6, which has implications on the physical processes underlying the relation.
A compilation of black hole limits for globular clusters\cite{lut13} concludes that the slope for those is closer to 2.3.
This rather shallower slope would imply the relation is defined by a process different to that in galaxies, perhaps suggesting that many of these systems are the stripped nuclei of dwarf galaxies.

Currently, this issue is wide open, and is unlikely to be resolved by currently available facilities.
The problem is again the extreme crowding, which occurs in the centres of the star clusters, exactly where one needs measurements in order to distinguish scenarios with and without black holes.
Proper motions, rather than just line-of-sight velocity dispersions, are needed in order to measure and account for anisotropy, which can have a significant impact on the black hole mass derived.
As indicated in Fig.~\ref{fig:47tuc}, suitable measurements will only become possible with ELTs where spatial resolution can overcome the crowding.

\section{High Contrast Imaging}
\label{sec:hci}

The study of planets around other stars is one of the fundamental science drivers for the ELT\cite{kis11}.
For MICADO, the two top level goals in this respect are to exploit the large aperture of the ELT in order to achieve a meaningful contrast at very small inner working angles, and to learn about how to perform high contrast imaging on ELTs as a pathfinder for future dedicated instrumentation.

\subsection{Characteristics}

The astrophysical opportunities for high contrast imaging on ELTs are very exciting.
And so, while MICADO itself is not primarily a high contrast imager, this mode will be implemented to the limits possible without compromising the standard and astrometric imaging modes.
Its characteristics include:
\begin{itemize}
\item
This is one of the main drivers for a SCAO system, which is able to provide a high Strehl ratio on axis using a bright NGS (while, naturally, one also aims to achieve good performance on fainter stars too). 
A pyramid WFS is a natural choice because it can reach higher Strehl ratios than a Shack-Hartmann WFS for the same star magnitude; and it also yields better performance in terms of contrast close around the star because of the lower speckle noise\cite{ric16}.
\item
Phase masks will be available in both the focal plane and pupil plane.
The former will include a small inner working angle phase mask\cite{hub18} (akin to an annular groove phase mask, or vortex, but optimised for the ELT pupil) and a classical Lyot coronagraph\cite{per17}, both to be used with a Lyot stop in the cold pupil.
The latter will be a grating vector apodised phase plate\cite{ott14,ott17}.
Neutral density filters are also required, to avoid saturation during acquisition and calibration exposures.
Sparse aperture masks\cite{lac14} will also be available.
\item
Since the field of interest is close around the parent star, it is only necessary to read out the central detector.
An option to increase the frame rate by reducing the number of rows read out will be available.
\item
Pupil imaging is needed in order to ensure there is a detailed record of how the pupil appears for each observing block. 
This is expected to be a standard calibration mode in MICADO, since the ELT pupil changes from night to night, due to `missing' segments, and the different individual segment transmissions (resulting from the segment cleaning schedule).
\item
The control software will provide an option for pupil stabilisation (de-rotation), to enable angular differential imaging (ADI).
Since the performance of the focal plane mask depends critically on its alignment with the star, the QACITS algorithm\cite{hub17} will be used to provide re-centering feedback at $\sim$0.1\,Hz rate during observations.
\item
The pipeline is only required to process individual frames.
The reason is that the way in which multiple exposures are combined, or used, depends on the science goals, and the optimal procedure is expected to differ between programmes.
In addition, processing techniques are developing very rapidly, so that a fixed pipeline procedure is an unattractive option.
\end{itemize}

\subsection{Astrophysical Applications}

\begin{figure}
\begin{center}
\includegraphics[width=17cm]{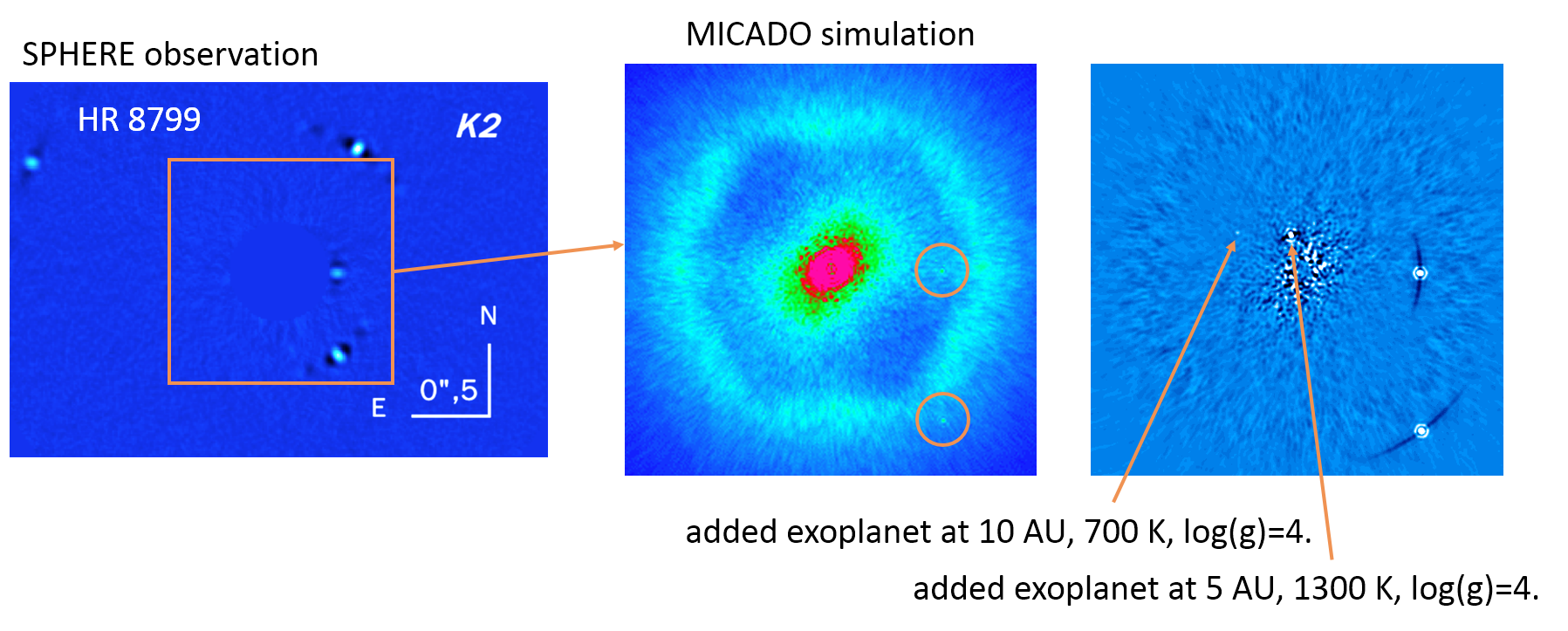}
\end{center}
\caption{Left: SPHERE image of HR\,8799 showing the four known planets\cite{zur16}. Centre: simulation of 30\,s integration with MICADO already reveals the two inner planets\cite{bau18}. The structure in the image arises from the optical configuration of the telescope and AO system, and its wavefront correction. Right: with basic processing of a series of ADI exposures, one is in principle able to detect other fainter and cooler planets at smaller radii.}
\label{fig:hr8799}
\end{figure}

Now that a large number of exoplanets are known, we are entering a phase driven by the need to characterise these planets, in particular the atmospheres of giant exoplanets.
Direct imaging of exoplanets provides an opportunity to do this through the use of intermediate band filters that cover molecular absorption bands, enabling one to distinguish models with different temperatures, surface gravities, and clouds.
The large aperture of the ELT offers a multiple gain for such work: the small inner working angle, the increased contrast between the PSF core and the speckles in the halo, and the elongation of the speckles when imaged through a broad or intermediate band (making them easier to distinguish from exoplanets).
As such, the focus for MICADO will be in terms of exoplanets at small orbital separations ($\sim$1\,AU) around nearby stars ($<$20\,pc); exoplanets at larger separations ($>$10\,AU) around nearby stars as well as more distant stars ($>$100\,pc), and the circumstellar disks from which they form.

Fig.~\ref{fig:hr8799}, which compares SPHERE observations of the planetary system around HR\,8799\cite{zur16} to simulations for MICADO\cite{hub18}, provides a glimpse of what it may be possible to achieve.
The central panel shows a coronagraphic simulation\cite{bau18} of how this system might appear with MICADO after adding 30\,s of exposures\cite{hub18}.
The elongated central region is the effect of the wind on the residual halo close around the suppressed PSF core\cite{bau18}.
The cleaned `control region' is clearly visible, as is its hexagonal boundary.
Even in this raw image, the inner 2 planets are already visible.
The right panel shows that, after basic processing, one is in principle able to see fainter planets closer in.
Two giant planets have been added, at 10\,AU and 5\,AU.
By imaging these through various intermediate band filters, one can estimate their temperature (700\,K and 1300\,K respectively, modelled using Exo-REM\cite{cha18}) due to their different molecular absorption properties.
This opens the very exciting potential that MICADO will be able to directly image planets for which a mass estimation is available from Gaia.

\section{Spectroscopy}
\label{sec:spectro}

The main rationale for spectroscopy in MICADO is to provide coverage of a wide wavelength range simultaneously at a resolution $R \sim 20000$, on faint compact or unresolved objects. 
In this sense it aims to emulate the success of X-shooter\cite{ver11}, while addressing a complementary role to the spatial resolution afforded by the integral field spectroscopic capabilities of HARMONI\cite{tha16,cla18}.

\subsection{Characteristics}

Spectroscopy in MICADO is a secondary mode, which means that its implementation must not compromise the primary imaging and astrometric imaging modes.
This has constrained the design choices available, but still led to a powerful capability, with the following characteristics:
\begin{itemize}
\item
The implementation of the spectroscopic mode in MICADO is optically smart, enabled by a single fixed cross-dispersing grating module that can be moved into the beam path by the main selection mechanism\cite{sch18}.
Order sorting filters are used to switch between wavelength ranges.
\item
Various slits are available in the focal plane mask. 
The narrowest, 16\,mas wide, is optimised for point sources.
Two slit lengths are available:
an off-centre 15\arcsec\ slit covers the 1.45--2.45\,$\mu$m (HK-band) and 1.15--1.35\,$\mu$m (J-band) wavelength ranges; 
a 3\arcsec\ slit is used for the 0.83--1.45\,$\mu$m (IzJ-band) range.
An additional set of slits, at slightly shifted positions, provides some flexibility to ensure that scientifically important spectral ranges do not fall across the gaps between detectors.
\item
Operationally, because the ADC is located in the cold pupil after the focal plane mask, the recommended alignment of the slit will be along the parallactic angle; with the orientation updated for each exposure.
The user, however, is not required to follow this scheme.
\item
The spectral resolution is around $R \sim 20000$ (15\,km\,s$^{-1}$) for point sources, and $R \sim 10000$ when integrated across the slit.
The difference arises because the slit is wider than the diffraction limit of the telescope, a necessary outcome of tolerancing and stability issues.
The key effect is that the spectral resolution depends on the spatial profile of the science target (and narrower than the background sky lines); and hence an image of the target in the slit is taken as part of the acquisition sequence, when centering the object.
\item
A metrology system is used to maintain optimal alignment of the science target in the slit.
The dominant issue here is the SCAO dichroic, which reflects the visible light to the WFS and transmits the near-infrared light.
This is a large optic in a large mechanism; and even though the NGS module is mounted stiffly to the cryostat, even very small drifts in its position could lead to a shift of the science target on the instrument focal plane that corresponds to a significant fraction of the 60\,$\mu$m (16\,mas) slit width.
The aim of the metrology system is to minimize these.
\item
The pipeline will rectify, calibrate, and combine the 2D spectral segments; and, for bright sources, extract a 1D spectrum.
A key design choice, to avoid too tight specifications on the mechanism repeatabilities, is that the pipeline should be able to handle small rotations and translations between the day-time calibrations and night-time science observations.
\end{itemize}

\subsection{Astrophysical Applications}

Spectroscopic simulations will be possible in the near future, as this mode of SimCADO has now been tested and is being released.
Science drivers include a wide range of topics:
the line-of-sight velocity dispersions of nearby galaxies, which can constrain orbit based models to derive black hole masses in galaxy nuclei, extending the parameter space to lower black hole masses as well as more distant galaxies;
measuring emission line spectra of early supernovae at $z \sim 1$--6; 
continuum absorption features in $z \sim 2$--3 early type galaxies, to measure stellar populations and dynamics; 
the metallicity, extinction, and dynamics of individual clumps in star forming galaxies at $z \sim 2$--6; 
the stellar types and 3D velocities of stars in dense stellar systems such as globular clusters and the Galactic Centre.

\section{Last Word}
\label{sec:lastword}

MICADO is the first light imaging camera for the ELT, and is being designed, constructed, and tested by a consortium of partners in Germany, France, The Netherlands, Austria, Italy, and at ESO.
The next project milestone is the Preliminary Design Review, which is scheduled for November 2018.

The instrument is optimised to operate with the multi-conjugate laser guide star adaptive optics system MAORY. 
The two consortia are also jointly developing a simple and robust single-conjugate natural guide star adaptive optics system. 
The observing modes offered by MICADO include: standard imaging, astrometric imaging, high contrast imaging, and slit spectroscopy. 
Observers will be assisted with dedicated observation preparation and data simulation tools, data processing pipelines, and PSF reconstruction.

\acknowledgments 
 
The consortium thanks the staff at the partner institutes and at ESO for continued support and enthusiasm for the project.

\bibliographystyle{spiebib} 

\end{document}